\documentclass[aps,prl,groupedaddress,reprint]{revtex4-2}
\usepackage{graphicx}
\usepackage{enumitem}
\usepackage{amsmath}

\newcommand{\NanoSecond}{$\;$ns}
\newcommand{\PicoSecond}{$\;$ps}
\newcommand{\FemtoSecond}{$\;$fs}
\newcommand{\CentiMeterSq}{$\;$cm$^2$}
\newcommand{\CentiMeter}{$\;$cm}
\newcommand{\MilliMeter}{$\;$mm}
\newcommand{\MicroMeter}{$\;\mu$m}
\newcommand{\NanoMeter}{$\;$nm}
\newcommand{\BarP}{$\;$bar}
\newcommand{\fps}{$\;$fps}
\newcommand{\MilliJoule}{$\;$mJ}
\newcommand{\Kelvin}{$\;$K}
\newcommand{\KiloKelvin}{$\;$kK}
\newcommand{\PerCC}{$\;$cm$^{-3}$}
\newcommand{\KelvinPerNanoSecond}{$\;$K$\;$ns$^{-1}$}
\newcommand{\eVPerNanoSecond}{$\;$eV$\;$ns$^{-1}$}
\newcommand{\eV}{$\;$eV}
\newcommand{\MicroMeterSqPerNanoSecond}{$\;\mu$m$^2\;$ns$^{-1}$}
\newcommand{\CentiMeterPerSecond}{$\;$cm$\;$s$^{-1}$}
\newcommand{\MicroMeterPerNanoSecond}{$\;\mu$m$\;$ns$^{-1}$}
\newcommand{\Intensity}{$\;$MW$\;$m$^{-2}\;$nm$^{-1}$}

\begin{document}


\title{Compressed Ultrafast Photography of Plasmas Formed from Laser Breakdown of Dense Gases Reveals that Internal Processes Dominate Evolution at Early Times}


\author{Peng Wang$^1$}
\author{Yogeshwar Nath Mishra$^{1,2,3}$}
\author{Seth Pree$^4$}
\author{Lihong V. Wang$^{1,*}$}
\author{Dag Hanstorp$^5$}
\author{John P. Koulakis$^6$}
\author{Daniels Krimans$^6$}
\author{Seth Putterman$^{6,*}$}
\affiliation{$^1$ Caltech Optical Imaging Laboratory, Andrew and Peggy Cherng Department of Medical Engineering, Department of Electrical Engineering, California Institute of Technology, 1200 East California Boulevard, Mail Code 138-78, Pasadena, CA 91125, USA \\$^2$ Science Division, Jet Propulsion Laboratory, California Institute of Technology, 4800 Oak Grove Drive, Pasadena, CA 91109, USA  \\$^3$ Present Address: Department of Physics, Indian Institute of Technology Jodhpur, Rajasthan 342030, India \\$^4$ Department of Applied Physics, California Institute of Technology, 1200 East California Boulevard, Mail Code 128-95, Pasadena, CA 91125, USA  \\$^5$ Department of Physics, University of Gotenburg, SE 412 96 Gothenburg, Sweden \\$^6$ Department of Physics and Astronomy, University of California Los Angeles, Los Angeles, California, USA \\$^*$ Corresponding authors: LVW@caltech.edu, puherman@ritva.physics.ucla.edu }


\date{\today}

\begin{abstract}
Compressed ultrafast photography (CUP) is applied to laser breakdown in argon and xenon under pressures up to 40\BarP\ to obtain 2D images of the plasma dynamics of single events with a spatial resolution of $250 \times 100$ pixels and an equivalent frame rate of 500 GHz.  Light emission as a function of position and time is measured through red, green, blue, and broad-band filters. The spatially encoded and temporally sheared image normally used in CUP is now enhanced by the introduction of a constraint given by a spatially integrated and temporally sheared unencoded signal. The data yield insights into the temperature, opacity, the plasma formation process, and heat flow within the plasma and to the surrounding ambient gas.  Contours of constant emission indicate that plasmas formed from sufficiently dense gas contract rather than expand despite having a temperature of a few eV. Plasmas formed from relatively low pressure gases such as 7\BarP\ argon can radiate with emissivity near unity.  Modeling transport and opacity as arising from inverse Bremsstrahlung requires a degree of ionization that strongly exceeds expectations based on Saha's equation even as customarily modified to include density and screening. According to this model, both electrons and ions are strongly coupled with a plasma coefficient $\gtrsim1$.  During the first few nanoseconds after formation, Stefan-Boltzmann radiation and thermal conduction to ambient gas are too weak to explain the observed cooling rates, suggesting that transport within the plasma dominates its evolution.  Yet, thermal conduction within the plasma itself is also small as indicated by the persistence of thermal inhomogeneities for far longer timescales.  The fact that plasma is isolated from the surroundings makes it an excellent system for the study of the equation of state and hydrodynamics of such dense plasmas via the systems and techniques described.
\end{abstract}


\maketitle

\begin{figure*}
 \includegraphics[width=0.98\linewidth]{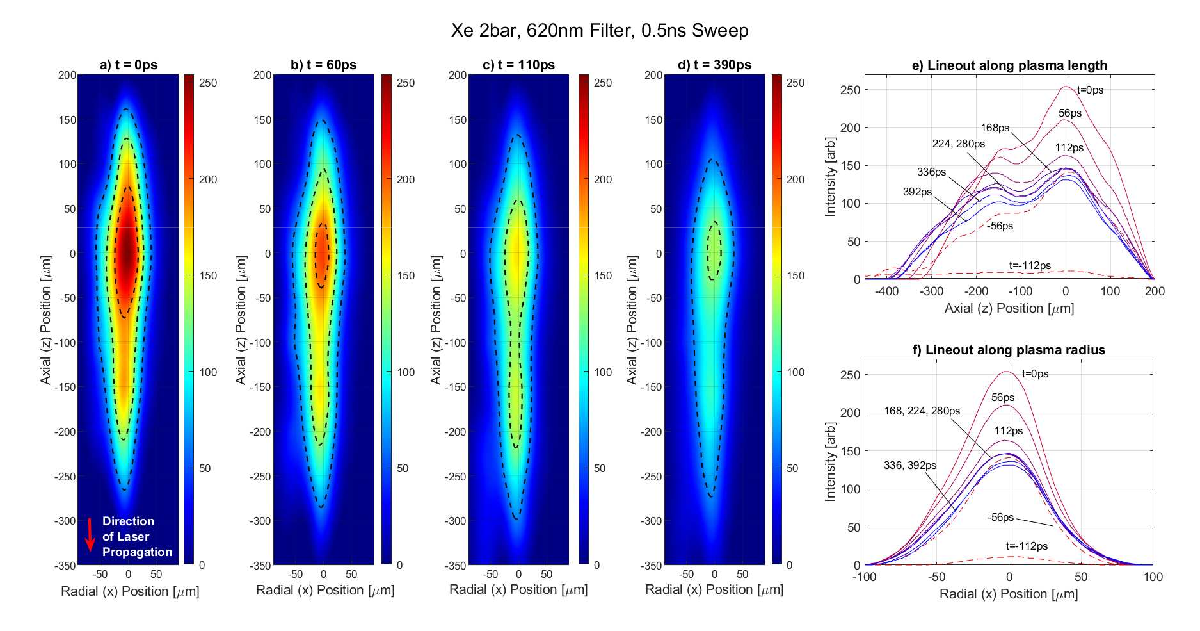}
\caption{\label{Figure1}a)-d) Four frames of the breakdown of 2\BarP\ xenon with a CUP sweep setting of 0.5\NanoSecond. Extracted frames are at t=0 (peak emission), 60, 110 and 390\PicoSecond.  A single-shot movie has 400 frames.  The intensity is false colored and in arbitrary units. Topographic lines of constant intensity are at 25\%, 50\%, and 75\% of the peak intensity of each frame.  e), f) Lineouts of the axial $z$ and radial $x$-dependence of the emission during the first 400\PicoSecond.  The lines cross the region of maximum emission intensity.  This data is from a single event. The fact that CUP has accurately captured multiple peaks in the axial direction is confirmed by traditional streak photos (Supplemental Material fig S7).}
\end{figure*}

\begin{figure}
\includegraphics[width=0.98\linewidth]{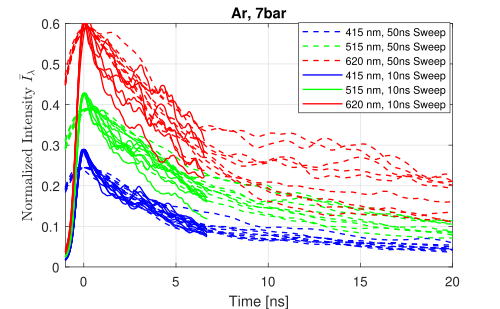}
\caption{\label{NormalizedIntensityFig}The measured emission (units of \Intensity\ per steradian) from the brightest 5x5 pixels region has been scaled with $\lambda^5/2hc^2$ to yield a dimensionless emission. Each trace is a single shot through Red (620\NanoMeter), Green (515\NanoMeter), and Blue (415\NanoMeter) filters for 10\NanoSecond\ and 50\NanoSecond\ sweeps.}
\end{figure}

\begin{figure}
\includegraphics[width=0.98\linewidth]{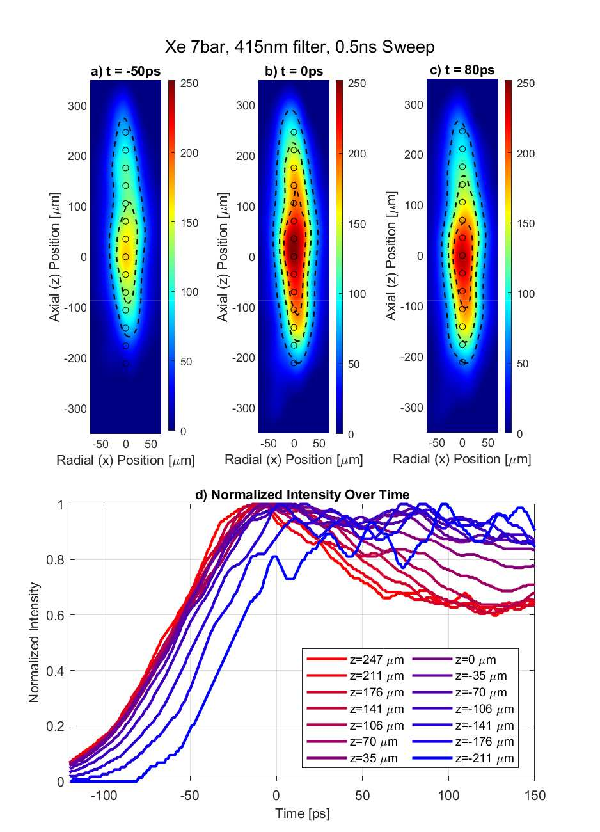}
\caption{\label{FastDynamicsFig}Rise of plasma emission as a function of position. Panels a), b), and c) are frames at $t=-50$, 0, and 80\PicoSecond\ respectively, where $t=0$ is the time of peak emission.  Circles in the frames indicate the positions where the light emission is plotted over time in panel d).  Emission in d) is normalized by the peak of each curve.}
\end{figure}

\begin{figure}
\includegraphics[width=0.98\linewidth]{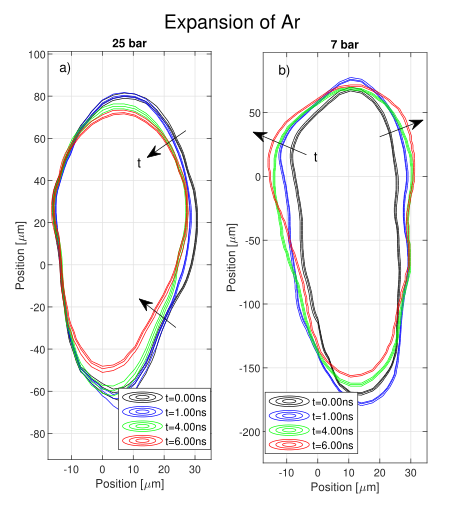}
\caption{\label{ExpansionFigure}Expansion of the breakdown plasma resolved by single shot CUP.  The evolution of 25\BarP\ argon a) and 7\BarP\ argon b).  Contours are drawn at $50\pm1$\% of the maximum intensity at various times.  The relative position of the contours indicates the expansion of the plasma.  For plasmas formed at high density, the contours steepen and the plasma contracts, whereas at 7\BarP\ the plasma expands.}
\end{figure}

In a strongly coupled plasma (SCP) the electrostatic potential energy is comparable to or greater than the thermal kinetic energy so that the fundamental plasma parameter $\Gamma=e^2/a k T\gtrsim 1$, where $e$ is the fundamental unit of charge, $T$ is the temperature, $ax^{1/3}=a_0=(3/4\pi n_0)^{1/3}$, $n_0$ is the atomic density and $x\leq1$ is the degree of ionization.  This regime is important for the study of astrophysical objects (e.g., white dwarf stars \cite{1968ApJ,whitedwarfnature}), dusty plasmas \cite{RevModPhys.81.1353}, ultracold atom/ion trapping \cite{KILLIAN200777,doi:10.1126/science.1130556}, ultrafast laser breakdown \cite{WOS:000499141000048,WOS:000341309300004}, photoelectron sources \cite{photoelectronheating} and inertial confinement fusion \cite{InertialFusion}. It should be contrasted with, say, Tokamak plasmas where the density is low and the temperature is high so that $\Gamma\ll1$ and where the constituent particles are weakly interacting similar to those in an ideal gas. On the other hand, for example, warm dense matter is a strongly coupled plasma which is described by Saumon et al. \cite{Saumon} as a regime where ``all the physics is important.''Despite numerous studies, controlling and measuring strongly coupled plasmas is a challenge that is here addressed with the application of Compressed Ultrafast Photography (CUP, \cite{CUP,Wang2024}) to the laser breakdown of a dense gas which we believe shows strong coupling.  Data generated by this technique show that on the nanosecond time scale the breakdown plasma is only weakly coupled to the environment so that its intrinsic thermohydrodynamic properties can be probed.

Due to the long-range nature of the Coulomb forces and the resulting correlation between particles, the macroscopic response of a SCP is also expected to be long-ranged \cite{PhysRevA.8.3096,PhysRevE.92.013107,Baalrud,10.1063/5.0194352} as compared to Navier-Stokes hydrodynamics.  So a line of attack for investigating SCP is to image their macroscopic motions. This is achieved by using CUP to capture the time-resolved, two-dimensional evolution of a SCP that is created by laser breakdown of dense argon and xenon. Example data that can be obtained with this technique are shown in Figures \ref{Figure1}-\ref{ExpansionFigure}.  Figure \ref{Figure1} shows the evolution of a plasma formed from laser breakdown of 2\BarP\ xenon as measured through a filter centered on 620\NanoMeter.  Breakdown is driven by a focused 0.9\MilliJoule, 220\FemtoSecond\ long pulse of 800\NanoMeter\ light using a windowed pressure cell \cite{WOS:000341309300004}.  Figure \ref{NormalizedIntensityFig} shows the intensity of emission from the bright central region of breakdown of 7\BarP\ argon as taken through 620\NanoMeter\ (``Red''), 515\NanoMeter\ (``Green''), and 415\NanoMeter\ (``Blue'') filters. The plotted emission intensity per steradian has been scaled to $\bar{I}_\lambda=I_\lambda\lambda^5/2hc^2$. Figure \ref{FastDynamicsFig} shows the rise of the plasma emission for different positions along the plasma length.   Since the curves at various positions have the same shape, we interpret the measured rise time as being limited by the system temporal resolution.  None-the-less, the time delay between curves at different positions gives information about the plasma formation processes.  Figure \ref{ExpansionFigure} shows the evolution of contours of constant emission that are normalized to the instantaneous peak intensity  for 7 and 25 bar argon.   These frames and intensity curves come from single shot data acquisitions and are not the result of patching together separate pump/probe acquisitions. The equivalent frame rate can reach 500 billion fps though we estimate the resolution to be 30\PicoSecond. 

Analysis of the spectral components indicates that argon and xenon at pressures above 7 bar emit with emissivity close to unity and are opaque.  At these atomic densities, the experimentally measured opacity requires a degree of ionization $x\gtrsim1$ assuming that opacity arises from inverse Bremsstrahlung. Combined with our measurement of temperature, we find that $\Gamma$ is order unity for both electrons and ions.  These systems can therefore be characterized as strongly coupled plasmas.

The goal of this communication is to describe the experiment which acquires the above data and the theoretical analysis which leads us to conclude that,
\begin{enumerate}
\item Stefan-Boltzmann radiation from the argon plasma is small compared to the measured changes in the energy of the plasma. 
\item Thermal conduction of energy to the surrounding ambient gas at 300\Kelvin\ is also small compared to the observed changes in energy.
\end{enumerate}
It follows that the response is dominated by the internal thermohydrodynamics of a strongly coupled plasma.  Furthermore, the plasma emission is not uniform and displays hot and cold spots.  Rather than evening out, with hot spots getting cooler and cold spots warmer, the emission drops locally in a way that its profile maintains its shape during the cooling.  This persistence of inhomogeneities within the plasma implies that the
\begin{enumerate}
	\setcounter{enumi}{2}
	\item Electron thermal conductivity inside the plasma is small compared to the timescale of the other transport processes.
	\item Local processes dominate thermal equilibration between species. If $\tau_{e-th}$ is the time scale over which the temperature of electrons and ions equilibrate, we arrive at the same conclusion as Bataller \cite{WOS:000341309300004} and interpret the rapid drop in emission at early times $t<\tau_{e-th}\lesssim1$\NanoSecond\ for the higher pressures as arising from electron-ion equilibration \cite{WOS:000341309300004,PhysRevLett.101.135001}.
\end{enumerate}
As the plasma cools and the electrons and ions equilibrate, some electrons will recombine and impart their ionization energy back to the electron gas causing recombination heating \cite{10.1063/1.323257,ABDELHAMEED2003271}. But,
\begin{enumerate}
	\setcounter{enumi}{4}
	\item Signatures of recombination heating are not observed.  
\end{enumerate}
We speculate as to whether the effects of screening \cite{PhysRevLett.111.234301,PhysRevLett.117.085001} are much greater than is theoretically proposed \cite{1966ApJ144.1203S,CROWLEY201484} for the parameter space realized in these experiments.

Certain regimes of SCP are predicted to have a tensile strength (negative pressure) \cite{10.1063/1.1727895}.  Laser breakdown in lower pressure (density) gas immediately expands radially, consistent with ideal gas dynamics and positive pressure plasma.  At higher pressure,
\begin{enumerate}
	\setcounter{enumi}{5}
	\item The plasmas display a lack of expansion, if not contraction, indicative of strong non-ideal interactions.
\end{enumerate}
This suggests that these systems might be a route to investigating the hydrodynamic consequences of the tensile strength of SCPs. 

\section{Experiment}
\begin{figure*}
\includegraphics[width=0.98\linewidth]{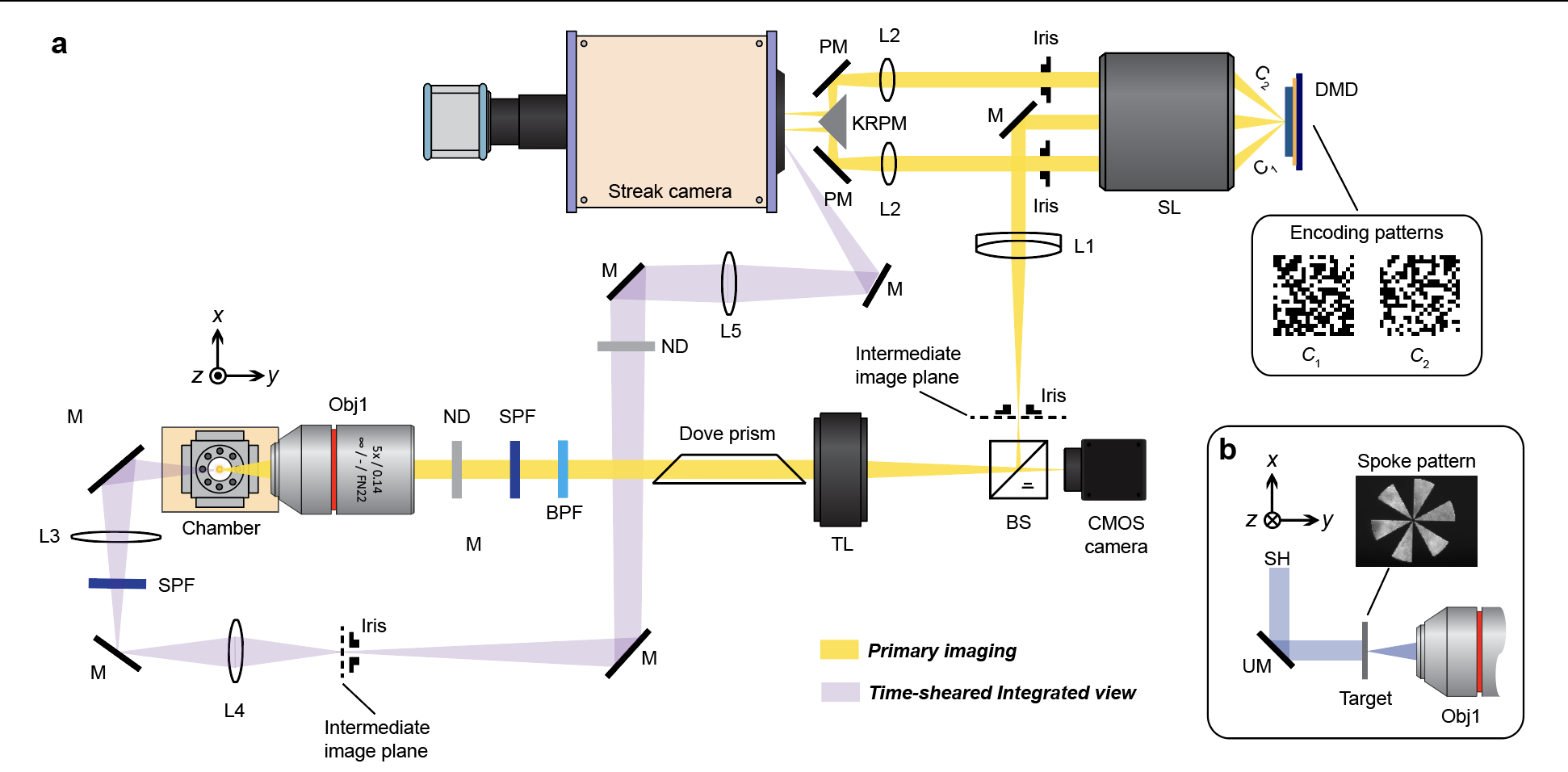}
\caption{\label{ExperimentalFig}Experimental CUP setup. a) Laser pulse propagating in the ‘z’ direction causes breakdown in the chamber. The yellow path provides spatially dependent emission to CMOS camera and Streak camera.  Prior to time shearing by the streak camera the signal passes through a coded aperture provided by a DMD. The intermediate image plane, DMD and streak camera photocathode are conjugate planes. b) The second harmonic of the fs laser is used to illuminate a test target for the purpose of measuring the response of the coded aperture acquisition and CUP inversion to a delta function, which forms the basis for selection of the optimal mask (see Supplemental Material \S 1.11).  Various optical elements such as band pass filters (BPF) and prisms (PM) are employed and listed with labels in Supplemental Material table S1.}
\end{figure*}

\begin{figure*}
\includegraphics[width=0.98\linewidth]{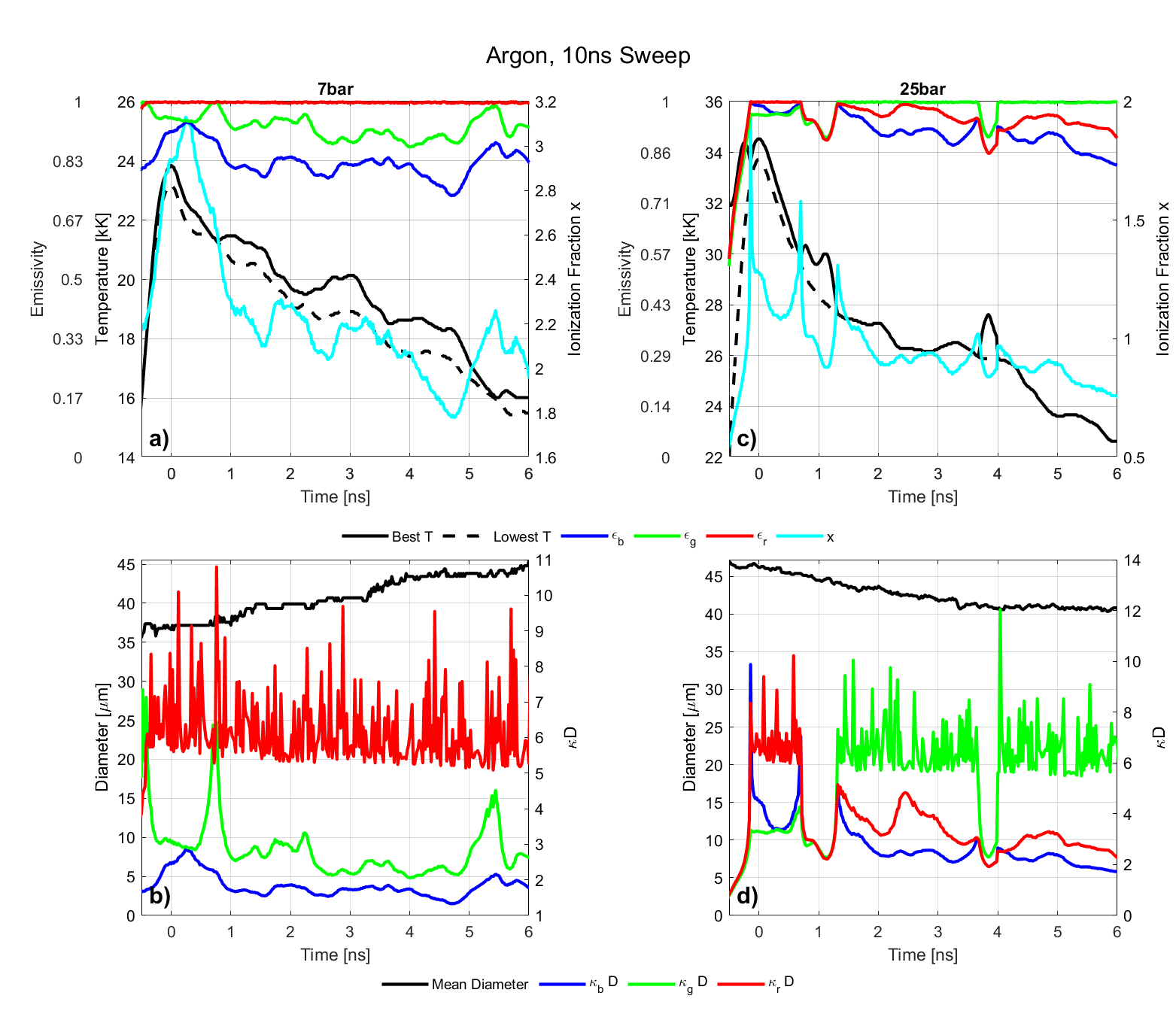}
\caption{\label{TemperatureAndIonization} Temperature, emissivity, ionization, mean diameter $D$, and optical thickness ($\kappa D$) for breakdown in 7\BarP\ (a,b) and 25\BarP\ (c,d) argon for 10\NanoSecond\ sweep.  Temperature is based upon emission from the central region of the plasma. Emissivities  $\epsilon_r$, $\epsilon_g$, $\epsilon_b$ are displayed for the best fit temperature. The dashed line is the minimum temperature at which the separate emissivities are all $\leq1$. The degree of ionization is determined from the lowest emissivity which is generally the blue curve.  The singularity in transport properties which accompanies emissivities close to unity leads to the spikes that can be seen in the curves for $\kappa D$, which is the photon mean free path.}
\end{figure*}

The experiment (figure \ref{ExperimentalFig}) uses a customized high-pressure chamber system built by Bataller \cite{WOS:000341309300004,Bataller:19} which supplies high-purity noble gas with adjustable and accurate pressure. A Ti:Sapphire femtosecond laser (Coherent) centered at 800\NanoMeter\ passes through a 60\MilliMeter\ focal length lens to induce breakdown at the center of the gas chamber, where it arrives with an energy of 0.9\MilliJoule.  The laser beam propagates in the z-direction, is linearly polarized in the x-direction (see figure \ref{ExperimentalFig}) and has a pulse duration tunable from 60\FemtoSecond\ up to a few picoseconds.  In this work, a temporally chirped pulse with 220\FemtoSecond\ duration was optimized and chosen to generate a plasma with brightest emission (Supplemental Material figure S9). See the full characterization of the laser system in the Supplementary Material figures S5 and S6.  The imaging system observes the plasma through the high-transmission quartz window on one side of the chamber.  An infinity-corrected objective lens in conjunction with a tube lens forms an intermediate image, which is split into two copies by a beam splitter: one is recorded by a conventional CMOS camera (static view, s.v. for short) and the other is further processed to form the time-sheared dynamic view (d.v. for short).  In this latter beam path, a digital micro-mirror device (DMD) (Texas Instruments) encodes the image by displaying a computer-generated 2D static pseudo-random binary pattern.  Thanks to the modulation mechanism of tilting mirrors, two complimentary encoded images are reflected by the DMD in two separate directions.  They are subsequently relayed and rerouted to a streak camera (Hamamatsu) side by side without overlap.  Inside the streak camera, temporal shearing of the frames in the dynamic scene achieves from 1 million\fps\ to 1 trillion\fps.  Signal synchronization between the laser and the streak camera is critical for the acquisition of any transient event.  Eventually, one single fs-laser pulse generates one single spatially-encoded temporally-sheared raw image of one complete evolution of 2D plasma dynamics on the streak camera.  Images are $250\times100$ pixels, and each pixel corresponds to a $3.5\times3.5$\MicroMeter\ plasma area.  See Supplemental Material for experimental setup and components.

Emission spectra are measured through narrow-band filters centered at 410\NanoMeter, 515\NanoMeter, and 620\NanoMeter, and a broad-band filter covering the entire visible light (350-650\NanoMeter).  The first three are extensively exploited to characterize the temperature and emissivity properties of blackbody radiation from plasma. 

Important to the interpretation of the data is an auxiliary view (purple path) introduced here which is taken from another window and projected directly to an unused region on the streak camera. It is spatially-integrated and time-sheared and provides a key constraint on image reconstruction. In addition, a small portion of the original laser pulse is frequency doubled to a 400\NanoMeter\ second harmonic (SH) pulse by a nonlinear BBO crystal. This SH pulse flood illuminates the chamber through its back window, which is essential in imaging and temperature calibration and described in Supplemental Material \S 1.4.

Recovering and de-compressing the 3D spatiotemporal $(x,z,t)$ dynamics of the observed event from the two simultaneously acquired 2D images: one from the CMOS camera and one from the streak camera, is a typical ill-conditioned inverse problem due to multiplexing and high compression, the solution of which can only be reached by resorting to regularization \cite{Wang2024}. Different from previous CUP modalities, the additional viewing channel introduced transforms the 3D data cube to a time sequence by 2D spatial integration. This added perspective towards the 3D domain presents a new angle of projection based on the theory of Radon transformation, potentially enforcing higher fidelity in reconstruction \cite{Wang2024}, and thus featuring a technological advancement from this work. The image forming pipeline, which is based on the assumption of data sparsity in the gradient domain, is discussed in the Supplemental Material \S 2.2. 

\section{Analysis}
Temperature and emissivity $\epsilon$ are determined by fitting the emission of the brightest 7x7 pixel region of the plasma in the 3 listed windows.  In general, the measured emission is $\bar{I}_\lambda=\epsilon_\lambda/\left[\exp(hc/\lambda kT)-1\right]$ where the emissivity $\epsilon_\lambda$ is unity for a Planck blackbody but in general depends on wavelength.  One restriction on $T$ is that $ \epsilon_\lambda\leq1$.  In other words, there is a minimum $T$ at which one of the colors has $\epsilon=1$ with the other wavelength having $\epsilon_\lambda<1$.  This value of $T$ is plotted in figure \ref{TemperatureAndIonization} as a lower bound.  Motivated by the gray body approximation where $\epsilon_\lambda$ is independent of wavelength, we choose a higher temperature which minimizes the spread in $\epsilon_\lambda$.  This temperature along with the various $\epsilon_\lambda$ is also plotted in figure \ref{TemperatureAndIonization}.

\begin{figure*}
\includegraphics[width=0.98\linewidth]{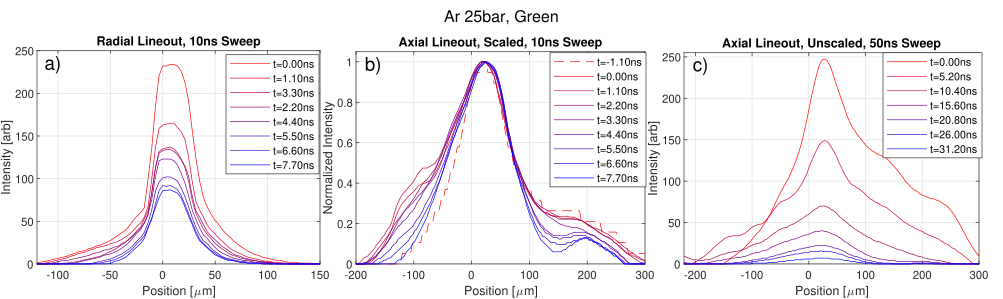}
\caption{\label{RadialAxialLineouts} Emission of a single shot along a lineout crossing the brightest pixels for argon 25\BarP\ acquired through the green filter.  a) is a radial lineout for a 10\NanoSecond\ sweep, b) is an axial lineout for a 10\NanoSecond\ sweep that is scaled to the peak value, and c) is an axial lineout for a 50\NanoSecond\ sweep.  The rapid drop in emission in (a) during the first nanosecond is due to local equilibration of electrons, ions, and neutrals.  Notice that the central peak in (b) maintains its curvature for at least 10\NanoSecond, which indicates the low level of thermal conduction within the plasma.}
\end{figure*}

A key parameter is the degree of ionization $x$. It determines the kinetic and ionization energy of a gas consisting of ions, free electrons, and neutral atoms:
\begin{align}
E=\frac{3}{2}n_0k(T_i+xT)+n_0x\chi_0,
\label{EnergyWithIonization}
\end{align}
where $n_0$ is the number of atoms in the plasma, $T_i$ is the ion/neutral temperature, $\chi_0$ is the tabulated ionization energy, and we have allowed for the possibility that the electron temperature $T$ has not yet equilibrated with the ions/neutrals.  To obtain $x$ from the spectral data  requires a higher level of theory as it involves an analysis of opacity and emissivity.  We start with the Nekrasov formula \cite{ADZERIKHO19751131} which connects the emissivity of a cylinder with diameter $D$ (estimated as the full-width-at-half-max emission) to the attenuation of light of wavelength $\lambda$ ($\kappa_\lambda$) in the plasma, $\epsilon_\lambda=1-\exp(-\kappa_\lambda D)$.  In view of the featureless spectra observed here and in previous experiments \cite{BatallerPHDThesis,WOS:000341309300004} we interpret the attenuation $\kappa$ in terms of inverse Bremsstrahlung which is free-free scattering.  In this case a Drude model \cite{PhysRevLett.113.024301} relates $\kappa_\lambda$ to a frequency dependent electron collision time  $\tau_\omega$ \cite{WOS:000341309300004}, 
\begin{align}
&\kappa_\omega=\frac{k \gamma}{\omega \tau_\omega}=x^n A \ln(1+\frac{B}{x^{n-3/2}}); \ \ \ \ \gamma=\frac{\omega_p^2}{\omega^2}=\frac{4\pi n_0 x e^2}{m \omega^2}\label{eq1}\\
&A=\frac{2}{\sqrt{6\pi}}\frac{\omega}{c}\frac{\omega^3_{p,1}}{\omega^3}\Gamma_1^{3/2}f(\bar{\omega}); \ \ B=\frac{0.7}{\sqrt{3}f(\bar{\omega})^{3/2}\Gamma_1^{3/2}}\\
&f(\bar{\omega})=\frac{1-\exp(-\bar{\omega})}{\bar{\omega}}; \ \ \bar{\omega}=\frac{\hbar\omega}{kT}\label{eq3}
\end{align}
where $m$ and $\Gamma$ are the mass and plasma coupling constant of an electron.  When $x>1$: $\Gamma=xe^2/a_0kT$; the subscript ``1'' means evaluated at $x=1$; for $x<1$, $n=2$ and for $x>1$, $n=3$.  Emission and degree of ionization are determined by the electron temperature $T$.  To obtain $x$ via opacity also requires the plasma diameter $D$ which is supplied by the radial lineout of intensity of emission as displayed for a single shot in Figure \ref{RadialAxialLineouts}a.  Values of $D$ and also $x$ are displayed in figure \ref{TemperatureAndIonization} as well.

\section{1. Blackbody Radiation is Small}
Blackbody radiation per unit surface area of the plasma is $\sigma T^4$ where $\sigma$ is the Stefan-Boltzmann constant.  In the approximation that the plasma is a cylinder with diameter $D$ and length $L$, the loss in energy due to radiation is $dE/dt|_r=-\pi DL\sigma T^4$.  The plasma energy will have contributions from the kinetic energy of ions, electrons and neutrals, as well as the potential energy of ionization.  We shall first consider 25\BarP\ argon at times after 2\NanoSecond\ so as to exclude the initial rapid drop in emission which is ascribed to electron-ion thermal equilibration.  Then the kinetic energy per unit volume is $(3/2) n_0(1+x) kT$ where $n_0$ is the density of the ambient, 300\Kelvin, 25\BarP\ gas.  Cooling due to blackbody radiation is $dT/dt|_r=-8\sigma T^4/[3kDn_0(1+x)]$.  Using $n_0=6.2\times10^{20}$\PerCC\ and $D=43$\MicroMeter\ yields cooling at a rate of 110\KelvinPerNanoSecond\ for parameters of $x=.9$ and $T=27$\KiloKelvin\ that apply at 2\NanoSecond.  A fit to the temperature in figure \ref{TemperatureAndIonization} gives,
\begin{equation}
T(t)=T(0)+A\left[e^{-t/\tau}-1\right]-B t,
\label{eq:TfitExp}
\end{equation}
where $T(0)=3.05$\eV, $A=0.5$\eV, $B=0.1$\eVPerNanoSecond, and $\tau=0.33$\NanoSecond\ so that the observed cooling rate is $B\simeq1100$\KelvinPerNanoSecond.  The $10\times$ higher cooling rate observed implies that radiative cooling is small and that other processes are more important.  Inclusion of the ionization potential $\chi_0=15.76$\eV\ would increase the heat capacity and further decrease the magnitude of the calculated cooling should $x$ be a decreasing function of $T$.

\section{2. Conduction to Ambient Gas}
Thermal conduction to the exterior ambient gas is limited by the thermal conductivity of ambient argon.  The maximum heat flow is realized under the assumption that the plasma temperature is constant with a step function initially at the boundary.  In this case the heat flow leads to a cooling given by \cite{Landau},
\begin{equation}
\frac{\Delta T}{T}=-\frac{5}{6(x+1)}\sqrt{\frac{\chi_Tt}{\pi D^2}},
\end{equation}
where $\chi_T\approx6\times10^{-4}$\MicroMeterSqPerNanoSecond\ is the thermal diffusivity of the ambient argon which surrounds the plasma and limits the heat flow to the environment.  We are using the same plasma model as for the discussion of radiative cooling.  For $t=4$\NanoSecond\ and $x=0.6$ this yields $\Delta T/T=3\times10^{-4}$ which again is small compared to the observed response of the plasma.

\section{Transport within the plasma}
Turning now to a discussion of transport time scales within the plasma we consider various cases for the collision cross-section $\sigma$:
\begin{enumerate}[label=\alph*)]
	\item $\sigma_{e,dilute}\approx A r_T^2 \ln{\Lambda}$ is the cross-section for an electron-electron or electron-ion collision in a low $\Gamma$ or hot plasma \cite{zeldovichsigmadilute}, where $r_T^2=\left(\frac{e^2}{kT}\right)^2$ is the impact parameter within which the collision is thermodynamically meaningful and $\Lambda=\delta_D/r_T$.	 Following Stanton and Murillo and Daligault \cite{PhysRevE.93.043203,PhysRevLett.101.135001}, $A=8\sqrt{2\pi}/3$.	 As an example, at 15,000\Kelvin\ and $x=1$ we have $r_T\sim1.1$\NanoMeter.  $\delta_D=\sqrt{kT/8\pi x n_0 e^2}$ is the Debye screening length, which is much greater than $r_T$ in the dilute limit.
	\item $\sigma_{e,D}\approx\frac{32\sqrt{2\pi}}{3}\delta^2_D=\frac{32\sqrt{2\pi}}{3}\frac{r_T^2}{6\Gamma^3}$ is the strong screening limit which applies when $\Gamma$ is large.  The expression will be used in the context of electron-electron transport processes where $n_e=x n_0$.  A numerical factor of 4 has been introduced so as to match the unifying transport formula of Stanton and Murillo \cite{PhysRevE.93.043203}.
	\item $\sigma_{e-int}=\frac{8\sqrt{2\pi}}{3}r_T^2 \ln\left(1+\frac{b_0}{\sqrt{x}}\right)$ which applies to the intermediate values of $1\lesssim \Gamma <3$ and $b$ is evaluated at $\bar{\omega}=0$.  This equation will be used for electron-electron and electron-ion collisions \cite{PhysRevLett.101.135001}.
	\item $\sigma_{e-0}=10^{-16}$\CentiMeterSq\ at 1\eV\ \cite{JFerch_1985} describes collisions of electrons with neutral argon atoms, with a collision time $\tau_{e-0}=1/v_{th}(1-x)n_0\sigma_{e-0}$ where $v_{th}=\sqrt{kT/m}$ and $m$ is the electron mass.  This expression applies when ionization is low.  
\end{enumerate}

\section{3. Thermal Conduction within Plasma}
Aside from conduction to the ambient gas, one can ask how heat moves to different locations within the plasma. In particular, how does thermal conduction within the plasma affect heat flow along the direction $z$ of the laser pulse where maximum inhomogeneity is expected.  Figure \ref{RadialAxialLineouts} shows lineouts through the region of peak emission at $r=0$ and $z=0$.  Panel a) displays a $\sim3\times$ drop in emmision during the first 10\NanoSecond.  Despite this, as shown in panel b), the curvature of the scaled emission profile barely changes with time, so that even within the plasma thermal transport has a small effect on the emission profile on the resolved timescales.  Although electrons are efficient transporters of heat \cite{zeldovichElThermalConduction} this observation is consistent with theory.

A general expression for the electron thermal diffusivity is $\chi_{T-e}=\frac{2\nu_{th}}{3\sqrt{\pi}\sigma n}$ where $n$ is the density of scattering sites.  The strongest electron thermal conduction is realized when the cross-section is given by case b). As an estimate, take $x\sim1$ and $T=30$\KiloKelvin, $n_0=6.2\times10^{20}$\PerCC, $\nu_{th}=6.9\times10^7$\CentiMeterPerSecond, $r_T=5.3\times10^{-8}$\CentiMeter\ so that $\Gamma\gtrsim0.7$, $\sigma=2.5\times10^{-14}$\CentiMeterSq\ to get $\chi_{T-e}=0.17$\MicroMeterSqPerNanoSecond.  The mean free path is $\ell\sim6.5\times10^{-8}$\CentiMeter.  Consider the diffusive spread in time of a Gaussian fit to the data. The width $w(t)$ at 80\% of the maximum should expand as $w^2(t)=w^2(0)+3.6\chi_{T-e} t$.  Using $w(0)\approx60$\MicroMeter\ yields an increase in width less than $1$\MicroMeter\ after $10$\NanoSecond.  This is consistent with experimental measurements which show that there is a very slight change in width, if not a contraction, during the first 7\NanoSecond.  We return to this issue in \S 7 below.   

\begin{figure}
\includegraphics[width=0.98\linewidth]{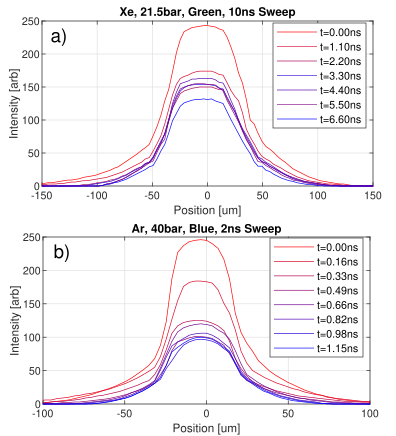}
\caption{\label{RapidEmissionDropArXe}Emission along a radial lineout through the brightest central pixels for a) xenon at 21.5\BarP\ and b) argon 40\BarP.  The rapid drop in emission during the first nanosecond is due to local equilibration of electrons with ions and neutrals.}
\end{figure}

\section{4. Electron-ion Equipartition of Energy}
The laser directly excites the electrons on a femtosecond timescale and so creates a situation where the ions have an initial temperature $T_i<T$.  The subsequent equilibration leads to a rapid drop in emission, which we interpret as a proxy for temperature, during the first nanosecond.  As we have argued, there is insufficient heat flow to the ambient gas, and thermal conduction within the plasma is low as well.  This leads one to interpret this drop as being due to electron-ion equipartition of energy.  After equilibration the drop in emission is slower.  This behavior is described with cross-section c) where collisions are partially screened \cite{PhysRevLett.101.135001, PhysRevLett.113.024301}, the parameter space for these experiments.  When applied to electron-ion collisions the time scale for thermal equipartition will be increased by the ion/electron mass ratio to $\tau_{e-th}=(M/m)\tau_{e-ion}$ where $\tau_{e-ion}=1/n_i\sigma_{e-int}v_{th}$, $M$ is the ion mass and $n_i=xn_0$ is the density of ions (taking here $x<1$).  This early time equilibration is qualitatively apparent in figure \ref{RapidEmissionDropArXe} and the best fit to the temperature (equation \eqref{eq:TfitExp}) yields a time scale of about 300\PicoSecond\ which should be compared with the 200\PicoSecond\ estimated with cross section c).

Ions can also be heated above their initial ambient temperature of $\sim300$\Kelvin\ by a process which is much faster than the collisional equilibration discussed above.  In our experiment, the laser pulse randomly rips electrons off the atoms on the femtosecond time scale.  In this initial plasma state the ions have a 2 particle correlation $g(r)=1$.  This function describes the probability of having two ions a distance $r$ apart.  However, $g(r)=1$, does not match the values in thermodynamic equilibrium under the given plasma coupling parameter, $\Gamma_i=e^2/akT_i$ for $x<1$ or, $\Gamma_i=Z^2e^2/a_0kT_i$ for $Z\geq 1$ where $Z$ is the degree of nuclear ionization.  For the case at hand of 25\BarP\ argon, $x=0.73$ at $t=0$, so $\Gamma_i\sim 65$.  This high value of $\Gamma_i$ means that the ions instantaneously constitute a very strongly coupled plasma which is out of the  equilibrium determined by the non-local Coulomb forces.  The approach to equilibrium results in a ``disorder induced heating'' (DIH) \cite{PhysRevLett.87.115003,doi:10.1126/science.1130556} which occurs on a picosecond time scale which is much shorter than $\tau_{e-th}$.  DIH would lessen the magnitude of the drop in emission that accompanies electron-ion equilibration.

According to Figure \ref{RadialAxialLineouts}b emission decreases more rapidly for $z\sim200$\MicroMeter\ with a decay rate that is approximately 2-3 times more rapid than at the peak.  We interpret this rapid decay as being due to lower ionization with a corresponding increased contribution from the neutral atoms. In this case the effective cross-section is described by d). This observation is made possible by the slow axial flow of heat. 

\section{5. Screening and Level of Ionization}

\begin{figure}
\includegraphics[width=0.98\linewidth]{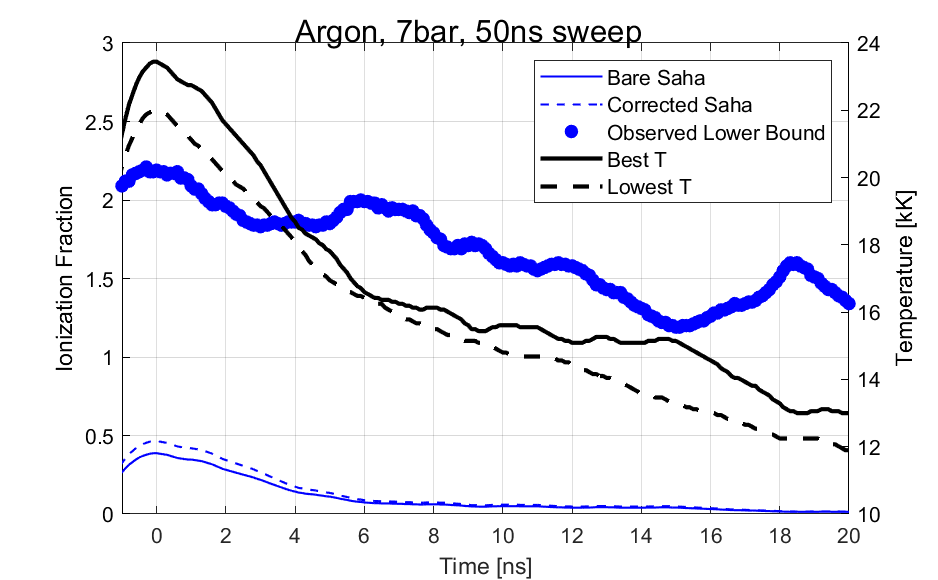}
\caption{\label{IonizationSaha} Temperature and ionization level of 7\BarP\ argon as compared to Saha's equation \eqref{sahaEquation}.  The thick blue line is the level of ionization as determined from emissivity and transport theory under the assumption that the photon mean free path is due to inverse Bremsstrahlung.  The thin solid blue curve is the level of ionization calculated with Saha's equation and the dashed blue curve is calculated with the first order screening effects included.  Both calculations are well below the lower bound, which suggests a larger screening effect or some other large source of continuum lowering.}
\end{figure}

Based upon emissivity which we interpret as arising from the scattering of light by free charges \cite{WOS:000341309300004,PhysRevLett.111.234301,PhysRevLett.106.234302,zeldovichElThermalConduction,Lee_2022,PhysRevLett.130.145103}, the deduced degree of ionization $x$ can be much larger than what follows from Saha's equation \cite{PhysRevLett.111.234301, PhysRevLett.117.085001, PhysRevLett.106.234302} even when leading order screening effects \cite{1966ApJ144.1203S,CROWLEY201484} are included.  Consider figure \ref{IonizationSaha}, which show the breakdown emission response of argon gas at 7\BarP\ taken under the same physical conditions as figure \ref{TemperatureAndIonization}, but for longer sweep or period of data acquisition.  Since the ionization energy for argon is $\chi_0=15.76$\eV\ and the temperature is 2\eV\ or less, one would expect that the plasma is single ionized at most.  This supposition is confirmed by Saha's equation of ionization,
\begin{equation}
\frac{x^2}{1-x}=\frac{2g}{n_0 \lambda^3_{deB}}e^{-\chi/kT},
\label{sahaEquation}
\end{equation}
where the thermal de Broglie wavelength is $\lambda_{deB}=h/\sqrt{2\pi mkT}$, $n_0=1.7\times10^{20}$\PerCC, $a_0=1.14$\NanoMeter\, and for argon the degree of degeneracy $g=6$.  According to Saha's equation, $x=0.4$ at 2\eV\ ($t=0$) and $x=.07$ at 1.4\eV\ which is the temperature at 7.0\NanoSecond.  The value of ionization deduced from opacity to blue light is much larger, being $x=2.2$ ($t=0$) and 2.0 ($t=7$\NanoSecond). As blue emissivity is lowest, we have used that value for this calculation.  The value of $x$ calculated from red and green emissivities is regarded as a lower bound as these values are close to one.

Saha's equation applies to an isolated atom and this version applies when only neutrals and singly ionized atoms are present so that doubly ionized atoms are inconsequential.  In a dense gas the ionization potential is lowered by screening which leads to an effective ionization potential \cite{1966ApJ144.1203S,CROWLEY201484},
\begin{equation}
\chi=\chi_0-\frac{kT}{2}\left((1+\Lambda)^{2/3}-1\right),
\label{CorrectedsahaEquation}
\end{equation}
where $\Lambda=\left(3x^{1/3}\Gamma_1\right)^{3/2}$ (for $x<1$).  Calculations of the level of ionization based on the bare \eqref{sahaEquation} and supplemented Saha equations \eqref{CorrectedsahaEquation} are plotted in solid and dashed blue curves in figure \ref{IonizationSaha}.   Both lie well below the minimum level of ionization observed and cannot account for the observed opacity based upon the attenuation of light being due to inverse Bremsstrahlung.  Either opacity is due to different dynamics, or some new processes \cite{PhysRevLett.106.234302} is contributing to a further lowering of the ionization energy.

\section{6. Lack of Recombination Heating}
Figures \ref{TemperatureAndIonization} and \ref{IonizationSaha} show the cooling of a 7\BarP\ argon plasma at a rate of 1100\KelvinPerNanoSecond\ between 0 and 5\NanoSecond.  Let us compare this to the average cooling due to blackbody radiation at the intermediate temperature of 20,000\Kelvin\ at 2\NanoSecond\ where $D=40$\MicroMeter.  This would lead to a drop in temperature of 90\KelvinPerNanoSecond\ when the heat capacity is that of an ideal gas.  This difference is accentuated when the possibility of recombination heating is evaluated.  From equation \eqref{EnergyWithIonization} a change in the degree of ionization $\Delta x=x(0)-x_f$ also would not only inhibit cooling but lead to heating as implied by energy balance:
\begin{equation}
\frac{3}{2}(1+x_f)kT_f=\frac{3}{2}(1+x(0))kT_0+\Delta x\chi_0+\frac{1}{N_0}\int_0^f\frac{dE}{dt}|_r dt\label{energyeq},
\end{equation} 
where subscripts $f$ and $0$ refer to final and initial states.  The last term is the radiated energy and for the purpose of developing this argument we are using the bare ionization potential.  Figure \ref{IonizationSaha} shows that the degree of ionization, as determined by opacity to blue light, decreases by $\Delta x=0.6$ during the interval $0<t<10$\NanoSecond.  Using $T(0$\NanoSecond$)=23,600$\Kelvin, and $\Delta x=0.6$ leads to $T_f=2.4 T(0)$ when $\chi_0=15.8$\eV.  This is a very large heating and there is no such signature in the data.  We propose that a larger-than-expected reduction in the ionization potential could account for both the high ionization and weak recombination heating.  

There are other sources of cooling. An expanding plasma does work against the exterior gas.  In view of the large pressure and temperature difference between the inside and outside of the plasma channel, this effect is small.  The hydrodynamic kinetic energy developed in an expanding plasma has also not been included in the energy balance.  Increases in this quantity during an expansion can lead to adiabatic cooling \cite{10.1063/1.2134768} which is discussed in the next section.  When the region of emission expands, we propose a means whereby screening can lead to cooling.  Emission from newly formed plasma requires local ionization which comes at the expense of recombination further inside the plasma, where the ionization energy is lower due to screening. 

\section{7. Lack of expansion}
\label{expansionsection}

A possible source of cooling is the adiabatic expansion of a gas.  Our experiment provides single-shot, time-resolved measurements of the plasma's two-dimensional macroscopic motion, enabling us to observe its changing shape over time as for instance indicated in figure \ref{ExpansionFigure}.  To what extent does this data give further insight into cooling and hydrodynamic motions?

In a simple model, laser breakdown creates an initial condition with a spatial discontinuity in pressure an temperature at constant mass density.  According to the Euler equations, such a contact discontinuity generates an outgoing shock wave as well as an expanding entropy (temperature) wave.  In an ideal gas in one dimension, such an entropy discontinuity initially propagates outward at the speed $u/2\gamma$, where $u(T,x)$ is the speed of sound on the hot side, and $\gamma$ is the adiabatic index.  The supplemental material shows this effect for the case where $T/T_0=100$ and $\gamma=5/3$ (see figure S13).

The initial state of the brakdown plasma also includes a discontinuity in ionization.  So the ideal gas kinetic pressure is $p=n_0(k T_i+x k T)$ is higher than used in the above model. For 25\BarP\ argon after electrons and ions have equilibrated, $T\sim 28$\KiloKelvin\ and $x\sim1$ leading to $p\sim4800$\BarP.  Using the adiabatic coefficient $\gamma=5/3$, a characteristic sound velocity is $u\sim\sqrt{\gamma p/n_0 M_{Ar}}\sim4.4$\MicroMeterPerNanoSecond.  For 7\BarP\ argon the characteristic temperature at 2\NanoSecond\ is 19\KiloKelvin\ with $x\sim2$ yielding $u\sim4.4$\MicroMeterPerNanoSecond. 

Figure \ref{ExpansionFigure} compares plots of contours of constant emissions for 7 and 25\BarP\ argon.  The expansion of the radius $D/2$ of the waist of emission from the breakdown of 7\BarP\ argon gas takes place at a speed of $0.7$\MicroMeterPerNanoSecond\ (see figure \ref{TemperatureAndIonization}) which is about a factor of 2 less than the $u/2\gamma$ expansion of a neutral ideal gas with the same sound velocity.  Figure \ref{ExpansionFigure}a shows the contours of constant emission for 25\BarP\ argon steepen, if anything, and do not propagate outward.  This applies in all directions.  For 7\BarP\ the plasma does not expand in the $z$-direction which on average is 7 times longer than the waist is wide.

Regarding the lack of expansion at 25\BarP, we note that one component plasmas can demonstrate a tensile strength \cite{10.1063/1.1727895}.  Whether these experiments are observing evidence for such effects will require more detailed modeling; in which we note that the ionization plotted in figure \ref{TemperatureAndIonization} should be regarded as a lower bound in view of the small deviation of emissivity from one.  Our time-resolved measurements suggest that we may be entering the regime where ion coupling and correlation effects play a significat role, which would provide a testing ground for various hydrodynamic \cite{10.1063/5.0194352, PhysRevE.92.013107} or kinetic \cite{PhysRevA.70.033416} models that incorporate these effects.

Self-similar solutions for the expansion of a plasma into a vacuum have been derived from hydrodynamics \cite{10.1063/1.2134768, 10.1063/1.1928247} or kinetic \cite{PhysRevLett.81.2691} equations for an ideal plasma. Although our initial plasma is not surrounded by vacuum, the ambient 25\BarP\ is small compared to the plasma pressure.  For Mora \cite{10.1063/1.2134768}, both the atomic density and the electron density are Gaussian. Due to the femtosecond nature of the breakdown which we study, the atomic density is best regarded as flat. However, a self-similar solution is also possible where the electron density is Gaussian and quasi-neutrality is maintained by a constant mass density which has a compensating space and time dependent ionization fraction.  In this model, the electrons expand at a velocity determined by the electron and not the ion mass. Such a rapid expansion is not seen.  

In general, the volume of the plasma contracts rather than expands.  For an ideal gas, this contributes to heating, rather than cooling.  Typical hydrodynamic equations cannot explain the macroscopic motion.

\section{Conclusion}
Here we point to the use of Compressed Ultrafast Photography in advancing the physics of a strongly coupled plasma for which the theory is incomplete. In this direction CUP has been used to resolve the time dependent 2-D behavior of a breakdown plasma formed from gases that are sufficiently dense that the plasma emissivity is close to unity. The key quantity measured is the plasma emission. Many physical processes contribute to the emission \cite{ModernRadiativeModeling}. In order to facilitate an application of the CUP data we have interpreted emission and absorption according to the free-free theory of plasma radiation and absorption (Bremsstrahlung and inverse Bremsstrahlung).  Plasma breakdown creates an ionized medium where the mean free path of light is due to its interaction with free electrons. Although this is in many ways the simplest model, we propose that it is in fact appropriate because detailed spectral measurements \cite{BatallerPHDThesis} are smooth and lack features. Its application points to a degree of ionization that can only be explained with a reduction in the effective ionization potential that exceeds existing theories.  Temperature is determined by a graybody fit and we find that an argon plasma formed at 25\BarP\ is an isolated system for about 10\NanoSecond\ during which time thermal conduction to the environment and radiative cooling are small. The two dimensional (radial and axial) response during electron ion equilibration is resolved.  Free expansion is observed at low but not high gas density.  As time progresses ionization decreases yet we do not discern signs of recombination heating.  We envision extension of this method to study 2D hydrodynamics especially during the time scales where $\Gamma>1$.  For 7\BarP\ argon at 5.5\NanoSecond, $T=16$\KiloKelvin, and $x=2$, yielding an ion coupling coefficient $\Gamma>3$.  For these parameters the Debye screening length $\delta_D=0.34$\NanoMeter\ and $\delta_D/a_0=0.3$ so that the ratio of Thomas Fermi to Debye screening lengths is $\delta_{TF}/\delta_D\simeq0.4$.  Figure \ref{Figure1} shows just a few frames from a movie with over 250 frames of a single event with a duration of 500\PicoSecond.  These systems are then test-beds for theories of equations of state and strongly coupled hydrodynamics, and sources for the unmasking of emergent properties of these systems.  It appears reasonable to apply this technique to obtain a single shot movie of processes that have been collected frame by frame using the pump/probe technique \cite{PhysRevA.95.013412} and to extend its use to x-ray-CUP of dynamical processes in dense systems \cite{ZHOU2024108508}.

\begin{acknowledgments}
This research has been funded by the AFOSR (No. FA9550-21- 1-0295) and by Caltech (Carver Mead New Adventures Grant). The views, opinions, and/or findings expressed are those of the authors and should not be interpreted as representing the official views or policies of the Department of Defense or the U.S. Government. We thank T. Chuna, Z. Johnson, M. S. Murillo, and L. G. Silvestri for many valuable discussions. We are indebted to C.C. Wu for calculations of the evolution of contact discontinuities.  YNM acknowledges Swedish Research Council Funding, Grant No. IPD2018-06783.
\end{acknowledgments}

\bibliography{CUP_Plasma}

\end{document}